\documentclass{PoS}
\usepackage{slashed}
\usepackage{graphics}
\usepackage{wrapfig}
\title{QCD phase diagram for finite imaginary chemical potential with HISQ fermions}

\ShortTitle{QCD phase diagram with imaginary chemical potential}

\author{\speaker{Jishnu Goswami}\\
	Fakult\"at f\"ur Physik, Universit\"at Bielefeld, D-33615 Bielefeld,
Germany\\
	E-mail:\email{jishnu@physik.uni-bielefeld.de}}

\author{Frithjof Karsch\\
	Fakult\"at f\"ur Physik, Universit\"at Bielefeld, D-33615 Bielefeld,
Germany\\
	Physics Department, Brookhaven National Laboratory, Upton, NY 11973, USA\\
	E-mail: \email{karsch@physik.uni-bielefeld.de, karsch@bnl.gov}}

\author{Anirban Lahiri\\
	Fakult\"at f\"ur Physik, Universit\"at Bielefeld, D-33615 Bielefeld,
Germany\\
	E-mail: \email{alahiri@physik.uni-bielefeld.de }}

\author{Christian Schmidt\\
	Fakult\"at f\"ur Physik, Universit\"at Bielefeld, D-33615 Bielefeld,
Germany\\
	E-mail: \email{schmidt@physik.uni-bielefeld.de }}

\abstract{
We present results from an ongoing study of the phase diagram
of (2+1)-flavor QCD using the HISQ action with smaller than physical
light quark masses in the
Roberge-Weiss (RW) plane on lattices with temporal extent $N_\tau=4$. We
find that the endpoint of the $1^{st}$ order RW transition line remains
$2^{nd}$ order at least down to light quark masses corresponding to Goldstone
pion masses of $55$~MeV. Furthermore, we show that the chiral condensate is
sensitive to the RW transition and may serve as a energy-like operator
characterizing universal behavior close to the RW endpoint.
}

\FullConference{The 36th Annual International Symposium on Lattice Field Theory - LATTICE2018\\
		22-28 July, 2018\\
		Michigan State University, East Lansing, Michigan, USA.}

\begin{document}

\section{Introduction}
The exploration of the structure of the phase diagram of strongly
interacting matter is an active field of research both in the experimental
and theoretical nuclear physics communities. For vanishing values of
the chemical potential it is well established that the
chiral transition is a crossover at physical values of the two light and 
the heavier strange quark masses.
However, in the limit of vanishing light quark masses the nature of the 
chiral transition is not yet established undoubtedly. In particular, it is
not settled whether the transition is $2^{nd}$ order 
or becomes $1^{st}$ order below a certain critical value of the light quark masses.
Earlier work from Pisarski and Wilczek \cite{Pisarski:1983ms} 
suggests that for two massless flavors
($N_f=2$) the chiral transition will be $2^{nd}$  order belonging to the $O(4)$
universality class if the axial $U(1)$ symmetry effectively gets restored
only at temperatures higher than that for 
$SU(2)_L\times SU(2)_R$ flavor chiral symmetry restoration. 
If the residual $U_A(1)$ breaking is small
already at the latter temperature the transition may become $1^{st}$ order.

\begin{figure}
	\centering
	\includegraphics[page=2,scale=0.22]{phase_diagram_col.pdf}\hspace*{-0.5cm}
	\includegraphics[page=1,scale=0.22]{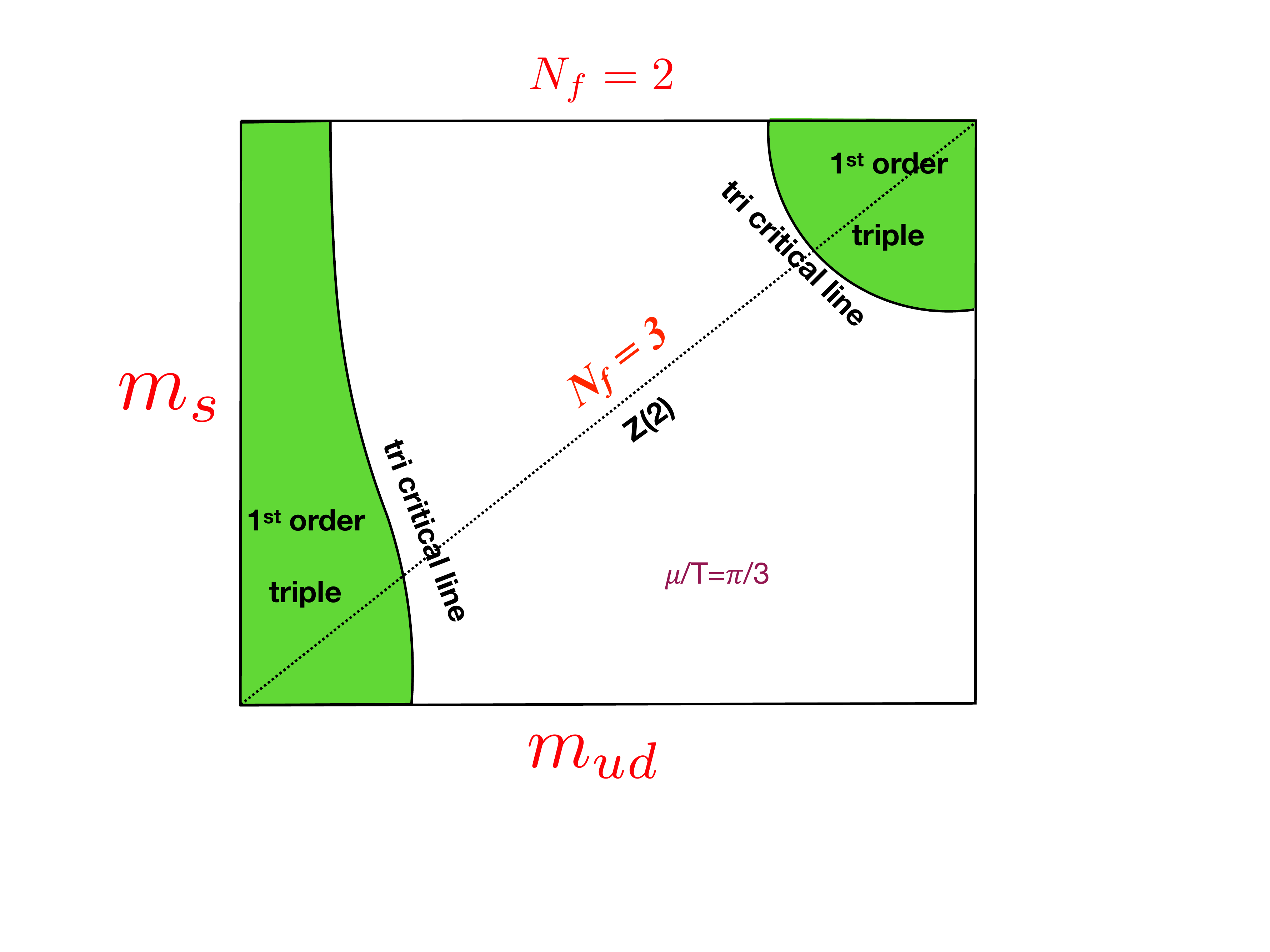}\vspace*{-0.8cm}
	\caption{Sketch of possible Coloumbia plots in the $\mu=0$ (left) and 
$i\mu/T=i\pi/3$ (right) planes.}
\label{Columbia}
\end{figure}

Studies with unimproved staggered fermions~\cite{deForcrand:2010he,Philipsen:2016hkv,Cuteri:2015qkq} show that the chiral transition is $1^{st}$ order below
a certain critical quark mass, which led to a possible version of the 
phase diagram in the light and strange quark mass plane (Columbia plot) as 
shown in Fig.~\ref{Columbia}~(left). However when using improved staggered
fermion actions, e.g. HISQ or stout, no $1^{st}$ order transitions is found even
for quite small quark masses, corresponding to Goldstone pion masses as small as 
50 MeV\cite{Ding:2018auz,Bazavov:2017xul}. This may imply that the chiral
phase transition in (2+1)-flavor QCD is $2^{nd}$ order in the limit of vanishing
light up and down quark masses. 

The calculations performed with unimproved staggered fermions~\cite{deForcrand:2010he,Philipsen:2016hkv,Cuteri:2015qkq} also suggest 
that the region of $1^{st}$ order chiral transitions increases in terms of 
critical quark or pion masses at non-zero imaginary values of the chemical 
potential (see Fig.~\ref{Columbia}~(right)), i.e. for negative $\mu^2$. 
Calculations performed at the largest $|i \mu/T|=\pi/3$ thus can set limits 
on $m_{\pi}^{cri}$ in the $\mu=0$ plane. 
Here we report on our ongoing studies of the 
QCD phase diagram with imaginary chemical potential and small values 
of the light quark masses, keeping the strange quark mass at its
physical value.
\vspace{-0.3cm}


\vspace{-0.2cm}
\section{Details of the Roberge-Weiss (RW) plane}
At imaginary values of the chemical potential, $i \mu$, the QCD partition 
function still has a real and positive fermion determinant. As the gauge
fields can always be transformed by globally multiplying all time-like
gauge field variables $U_{x,\hat{0}}$ with an element of the center of
the SU(3) group,
\begin{wrapfigure}{l}{0.47\textwidth}
\vspace*{0.5cm}
        \centering
        \includegraphics[scale=0.18]{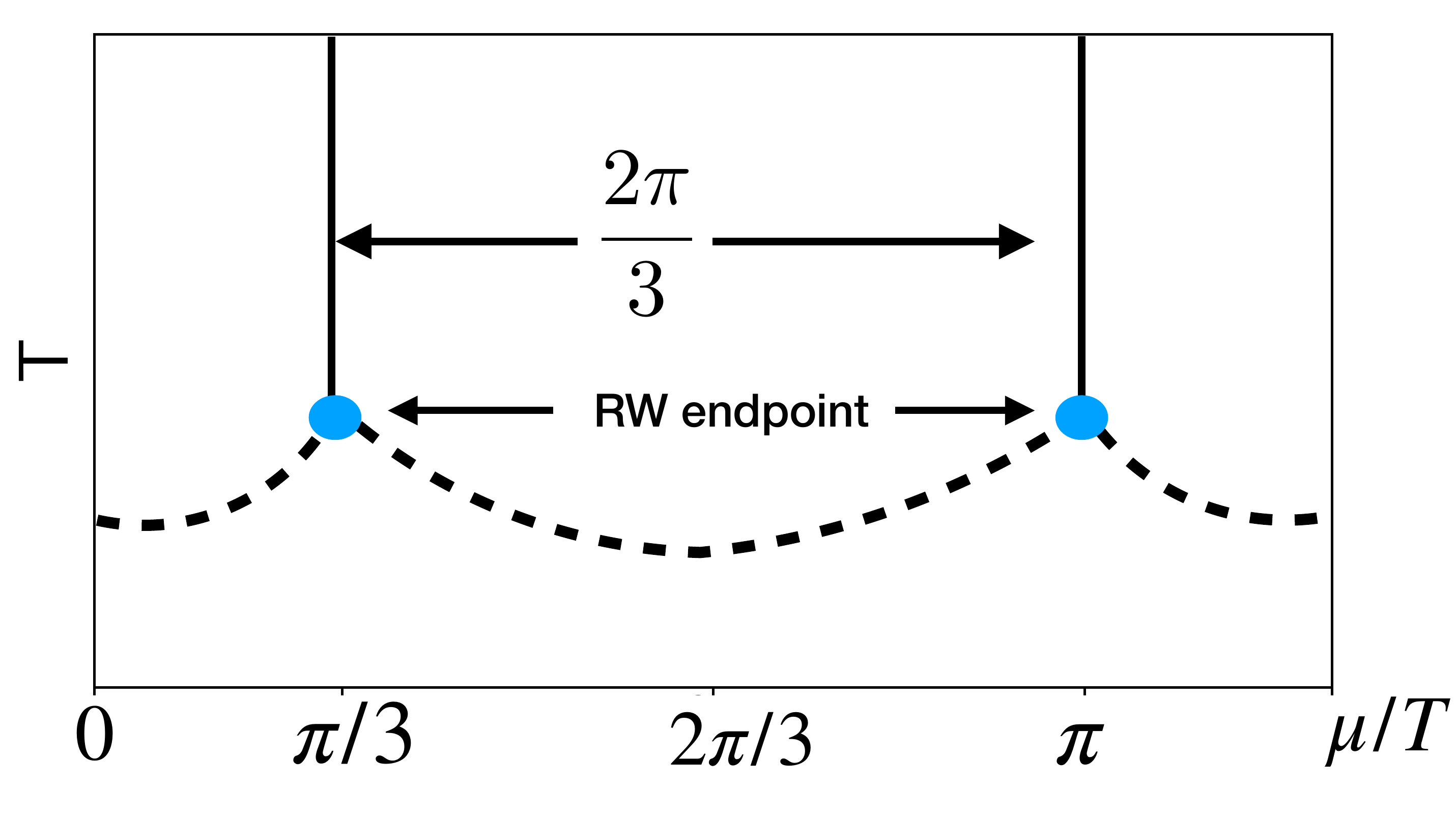}
\vspace*{-0.2cm}
        \caption{A possible phase diagram in the imaginary chemical potential
plane. Solid lines show first order phase transitions, dotted lines correspond
to crossover transitions and the blue points indicate $2^{nd}$ order phase
transition points.}
    \label{phase_imL}
\end{wrapfigure}

\noindent
the partition function is periodic under shifts $\mu/T\rightarrow \mu/T+2\pi/3$, 
{\it i.e.} $Z(\mu/T)=Z(\mu/T+2\pi/3)$. It has been pointed out by
Roberge and Weiss \cite{Roberge:1986mm} that this leads to a periodicity 
at $\mu/T=(2k+1) \pi/3$
and that a phase transition
may occur for these values of the imaginary chemical potential at high
temperature. These particular choices of the imaginary chemical potential
in a given Z(3) sector are known as the Roberge-Weiss (RW) planes.

A specific scenario for phase transitions in the RW plane, consistent with the 
results presented here, is shown in 
Fig.~\ref{phase_imL}. Alternative scenarios have been found in 
calculations with standard staggered and also Wilson fermions,
where the RW endpoint turns out to be a triple point for sufficiently
small values of the light quark masses and three first order transition lines
would emerge from this triple point.
For $N_f =2$ a $1^{st}$ order triple point 
has been found  on coarse lattices with temporal
extent $N_\tau =4$ in calculations with standard staggered fermions below 
$m_\pi^{cri}\sim 400$ MeV  and with the standard Wilson fermion action below
$m_\pi^{cri}\sim 930$ MeV. The latter critical mass shifted to 
$m_\pi^{cri}\sim680$ MeV for $N_\tau=6$. 
Results thus are found to be strongly dependent on the fermion 
discretization scheme and cut-off ($N_\tau$) 
\cite{deForcrand:2010he,Philipsen:2016hkv,Cuteri:2015qkq}. It therfore is not
too surprising that these results can change drastically when using
improved discretization schemes. In calculations with the stout action 
and for $N_f =2+1$ it is found that for physical pion mass the RW 
endpoint belongs to the Z(2) universality class. 
No $1^{st}$ order triple point has been found
at least for $m_{\pi} >50$ MeV~\cite{Bonati:2016pwz,Bonati:2018fvg}. In 
calculations using the 2-flavor HISQ action no clear-cut results on the 
order of the phase transition have been reported so far \cite{Wu:2018oed}.
In our calculations we use the (2+1)-flavor HISQ action together with an
${\cal O}(a^2)$ improved gauge action
to examine the nature of the RW-endpoint at smaller than physical values
of the light quark masses and physical value of the strange quark mass. 


\section{Ising universality and finite size scaling}

\begin{figure}
	\centering
	\includegraphics[scale=0.7]{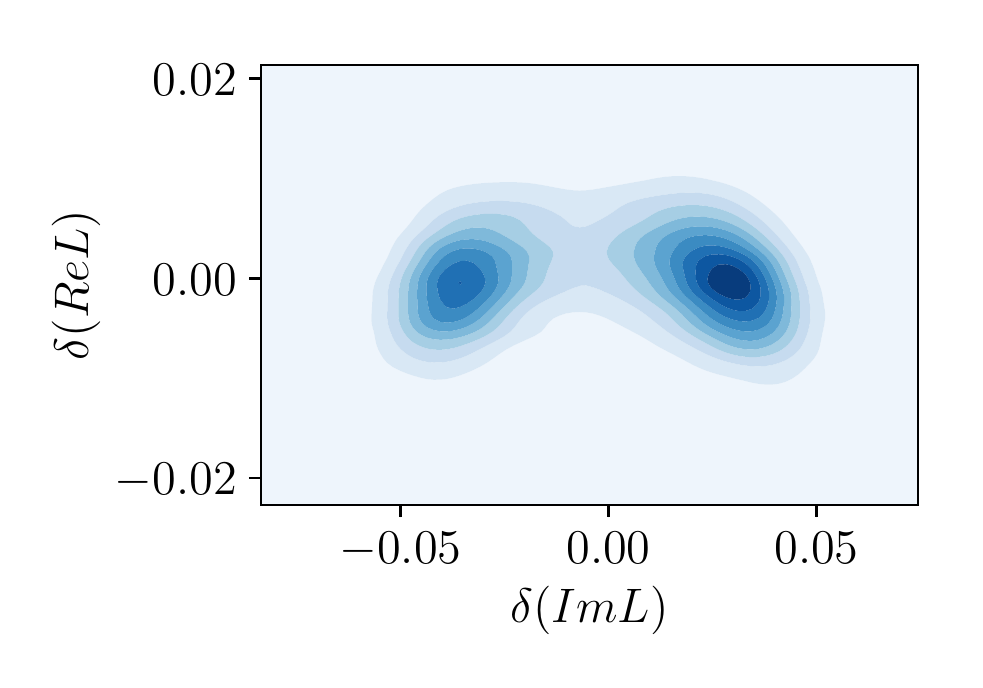}
	\includegraphics[scale=0.7]{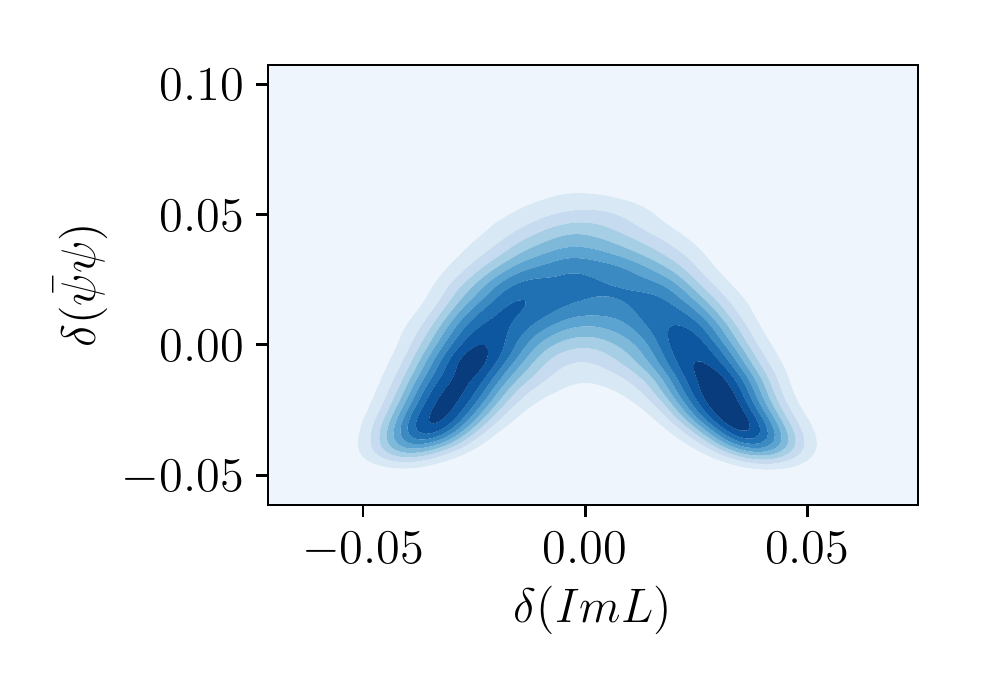}
\vspace{-0.2cm}
	\caption{Contour plot of $Re~L$ (left) and 
$\bar{\psi}\psi$ (right) versus $Im~L$ in the vicinity of 
the RW endpoint for $m_l/m_s =1/27$.}
\label{contour}
	\end{figure}

If the endpoint of the line of first order transitions in the RW plane
is second order, it will belong to the 3-d Z(2) universality class.
Near this critical point physics can be described by an effective Ising 
Hamiltonian, which characterizes the universal critical behavior of any system 
going through a Z(2) transition. We define
\begin{equation}
H_{eff}(t,h)=t\mathcal{E}+h\mathcal{M} \; ,
\end{equation}
where, $t$ and $h$ are temperature and external field like couplings, that
couple to an energy-like operator, $\mathcal{E}$ and a magnetization-like
operator $\mathcal{M}$.
Under Z(2) transformation, $\mathcal{E}\to\mathcal{E}$ and $\mathcal{M}\to-\mathcal{M}$, {\it i.e.} $\mathcal{E}$ remains invariant while $\mathcal{M}$ changes 
sign. Possible choices for these operators are the real and imaginary parts
of the Polyakov loop, $\mathcal{E}\sim Re~L$, $\mathcal{M} \sim Im~L$.
However, other choices are possible. For instance, the chiral condensate,
$\bar{\psi}\psi$,
is invariant under a Z(2) transformation that changes the imaginary part
of all Wilson loops. It thus may equally well serve as the energy-like
operator in an effective Ising Hamiltonian. We show contour plots of 
energy-like operators versus the magnetization like operator in 
Fig.~\ref{contour}. They present the characteristic features 
of ``banana-shaped'' contours known for Z(2) transitions 
\cite{Rummukainen:1998as}, which is even more obvious for $\bar{\psi}\psi$
than $Re~L$.

The logarithm of the QCD partition function defines the free
energy density, $f(T,\mu,V) = -\frac{T}{V} \ln Z(T,\mu,V)$, which in the 
vicinity of a critical point can be split into a singular and regular
part,
\begin{eqnarray}
f\simeq  b^{-d}f_s(b^{1/\nu} t/t_0, b^{\beta\delta/\nu} h/h_0, b^{-1}N_\sigma/l_0) + f_{r}(T,\mu,V) \; ,
\label{scaling}
\end{eqnarray}
where $f_s$ is a homogeneous function in terms of
the reduced temperature $t=(T-T_c)/T_c$, the symmetry breaking field 
$h=\mu/T-\pi/3$ and the volume $V=N_\sigma^3$. They all are
expressed in units of non-universal scale
parameters $t_0$, $h_0$ and $l_0$. Eq.~\ref{scaling} describes the universal 
critical behavior close to the critical point $(t,h)=(0,0)$ in the RW plane. 
From this one may derive the scaling functions for the order parameter 
$M= -\partial f(T,\mu,N_\sigma)/\partial h$ and its susceptibility
$\chi_h= -\partial^2 f(T,\mu,N_\sigma)/\partial h^2$. As the 
symmetry breaking field vanishes in the RW plane, $h=0$, 
we may set $b=N_\sigma/l_0$
to obtain the finite size scaling relations for $M$ and $\chi_h$,
\begin{eqnarray}
M &=& z_1 N_\sigma^{-\beta/\nu} f_G (z_0 tN_\sigma^{1/\nu}) + \; reg. \; ,\\
\chi_h &=& z_2 N_\sigma^{\gamma/\nu}~f_\chi(z_0 t N_\sigma^{1/\nu}) + \; reg. \; ,
\label{sus_scaling}
\end{eqnarray}
with constants $z_0$, $z_1$ and $z_2$ that are related to the scale parameters
$t_0$, $h_0$, and $l_0$.
We also adopted the conventional notation and normalization for the 
universal scaling functions of the order parameter and the susceptibility, 
$f_G$ and $f_\chi$, respectively \cite{Engels:2002fi}.


\section{Simulation details}
We have performed our calculations at imaginary chemical potential
in the RW plane using the HISQ action. This reduces ${\cal O}(a^2)$ 
cut-off effects in the staggered fermion discretization scheme and efficiently 
suppresses taste-changing interactions. This provides an improved 
approximation of continuum physics relative 
to that reached with standard staggered action at the same value of the cut-off.

The partition function for (2+1)-flavor QCD with two degenerate light quark 
masses ($m_l$), a strange quark mass ($m_s$)
and identical chemical potentials $\mu/T=\pi/3$ for all flavors
may be written as,
\begin{equation}
Z(T,\mu)=\int [\mathcal{D}U] det[M_{l}(i\mu)]^{1/2}det[M_s(i\mu)]^{1/4}\exp[-S_G] \; ,
\end{equation}
where, $M_q=D_{HISQ}(i\mu)+m_q$. For the gauge action, $S_G$, we use the tree 
level improved Symanzik action. The strange quark mass has been fixed
to its physical value and the light quark masses have been 
varied starting
at the physical value, $m_l/m_s=1/27$, towards chiral limit. The smallest
value used in our calculations, $m_l/m_s=1/160$, corresponds to a Goldstone
pion mass of about 55 MeV. Some further details on our simulation parameters 
are given in Table 1.
\begin{table}[hbt]
	\centering
	\begin{tabular}{|c|c|c|c|}
		\hline
		$N_\sigma$ & $N_\tau$ & $m_l/m_s$ & $m_{\pi}$(MeV)\\
		\hline
		8 & 4 & 1/27 &  135\\ 
		\hline
		12 & 4 & 1/27 &  135\\
		\hline
		16 & 4 & 1/27, 1/40, 1/60 &  135, 110, 90 \\ 
		\hline
		24 & 4 & 1/27, 1/40, 1/160 &  135, 110, 55 \\
		\hline
	\end{tabular}
	\caption{Details of the numerical simulation parameters}
\end{table}
We vary the temperature in the range, $T\sim T_c\pm 0.1T_c$. Generally we generated 20,000 trajectories per $T$ value away from $T_c$ and 80,000 trajectories 
near $T_c$. 


\section{Results}
\subsection{Scaling in the vicinity of the RW endpoint}
\begin{figure}
	\centering
	\includegraphics[scale=0.54]{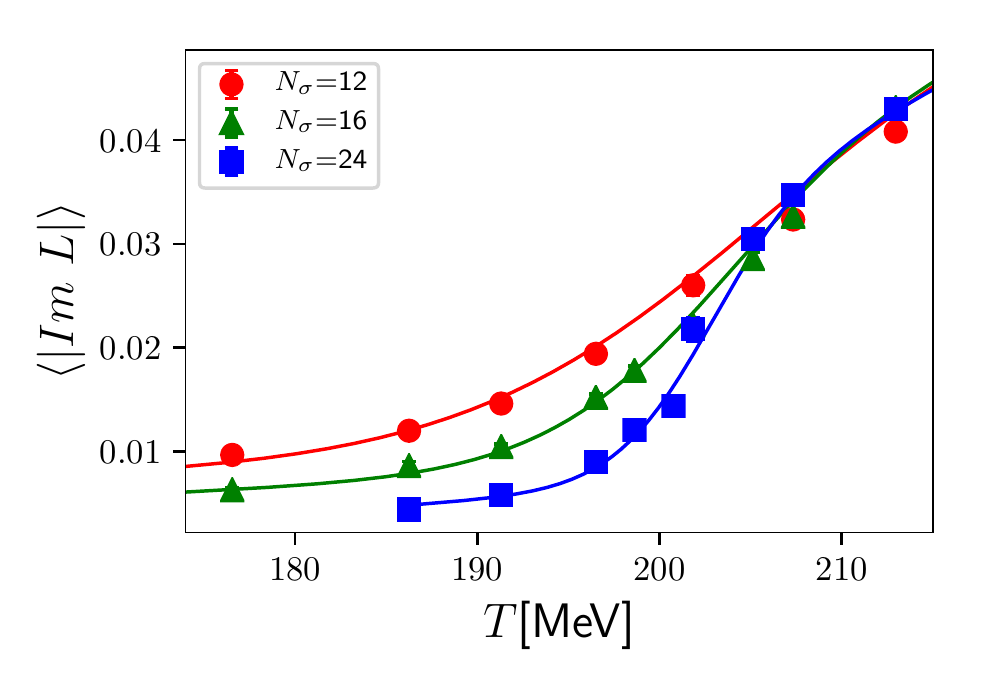}
	\includegraphics[scale=0.54]{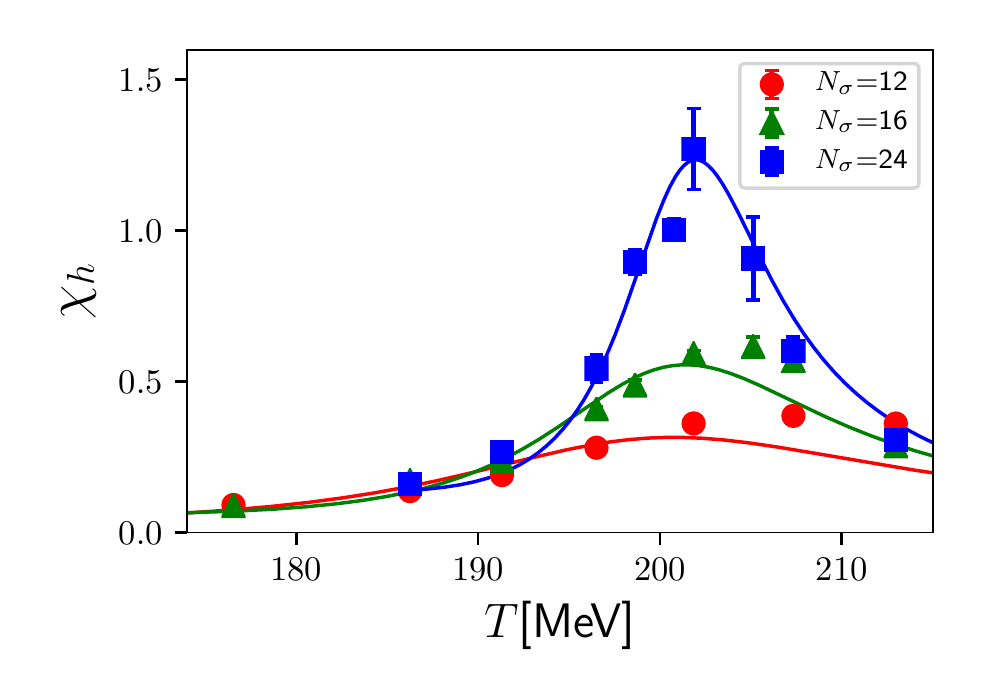}

\vspace{-0.2cm}
	\includegraphics[scale=0.54]{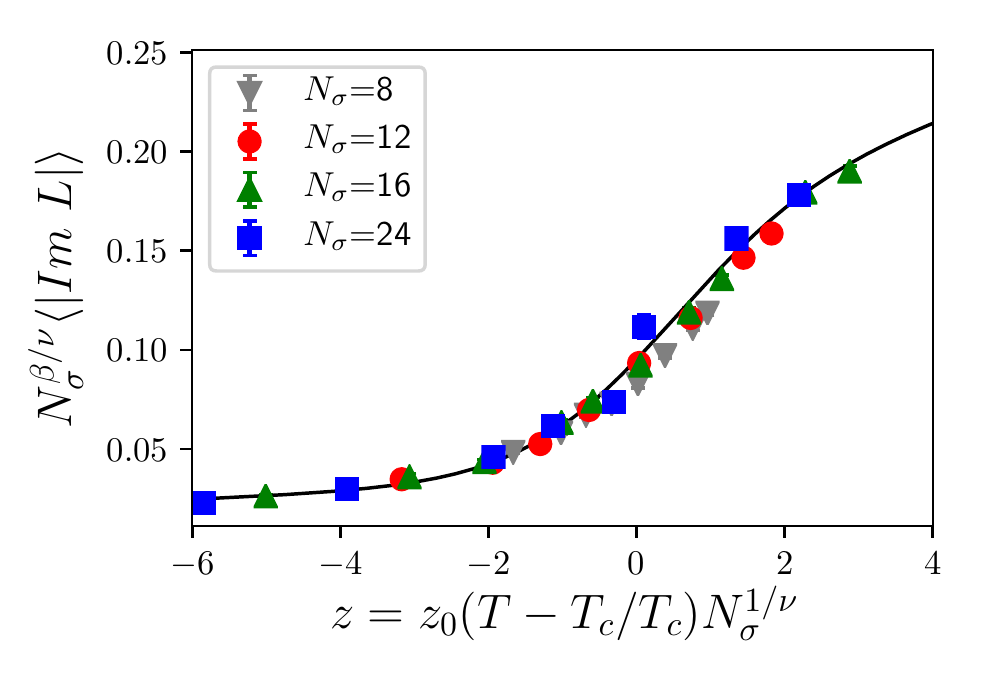}
	\includegraphics[scale=0.54]{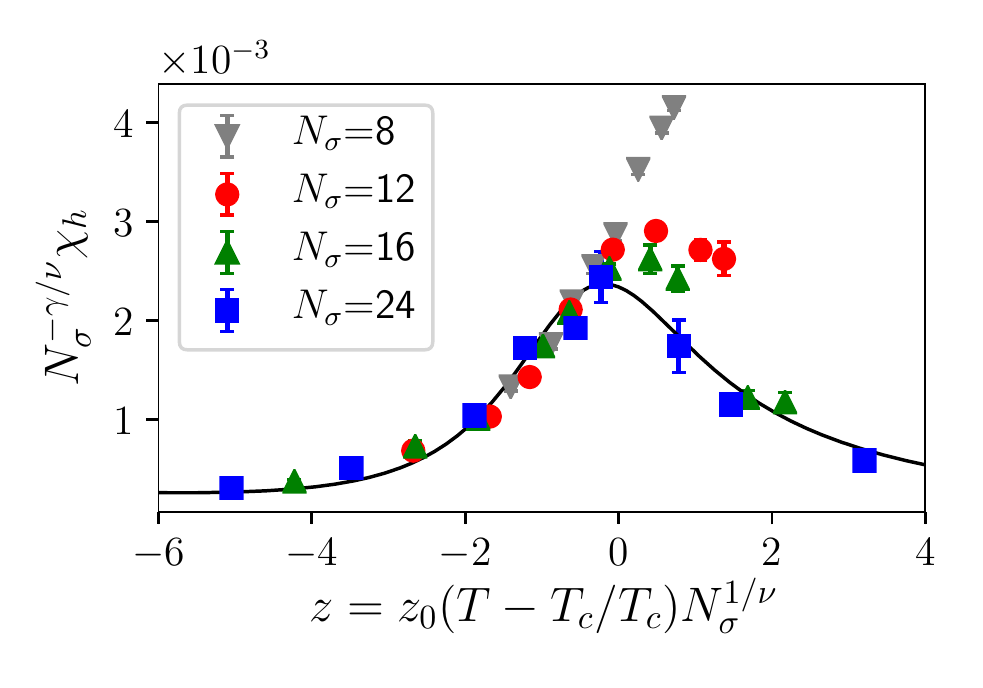}
    \caption{The order parameter $M$ (top, left) and the order parameter susceptibility $\chi_h$ (top, right) for $m_l/m_s=1/27$ and various lattice sizes.
Lines correspond to fits using the Z(2) finite size scaling functions. The
corresponding rescaled data and scaling functions are shown in the figures
on the bottom.}
   \label{op}
\end{figure}

By varying the light quark masses in our simulations we examined the 
behavior of the order parameter $M$ and the susceptibility $\chi_h$
as function of the spatial volume. Even at the smallest quark mass
value used in our simulations, $m_l/m_s=1/160$, we find that the 
peak of the susceptibilities at $T_c$ rises significantly slower 
than the $N_\sigma^3$
divergence that would be expected for a first order phase transition. 
We thus compared our results to the expected scaling behavior for a 
transition in the 3-d, Z(2) universality class. In Fig.~\ref{op} (top) 
we show results for the order parameter $M=\langle |Im L|\rangle$ (top, left)
and the order parameter susceptibility, $\chi_h$ (top, right),
calculated on lattices with spatial extent $N_\sigma =12,\ 16$, and $24$ and 
$m_l/m_s=1/27$. These results for $N_\sigma =16$, and $24$ have been fitted 
using the Z(2) finite size scaling functions. The rescaled data are shown in
Fig.~\ref{op} (bottom) and include also results for $N_\sigma =8$ for
comparison. As can be
seen for the order parameter possible contributions from regular terms that 
lead to deviations from the universal Z(2) scaling curve are visible for 
$N_\sigma =8$, but are small. For the susceptibility, however, scaling 
violations are large in the symmetry broken phase. This also has been
observed in simulations of the Ising model and, in addition to contributions
from regular terms, may indicate substantial contributions from corrections
to scaling \cite{Engels:2002fi}. 
Similar results we find for smaller values of the light quark masses.

\subsection{Chiral observables in the vicinity of the RW endpoint}

\begin{figure}
        \centering
        \includegraphics[scale=0.36]{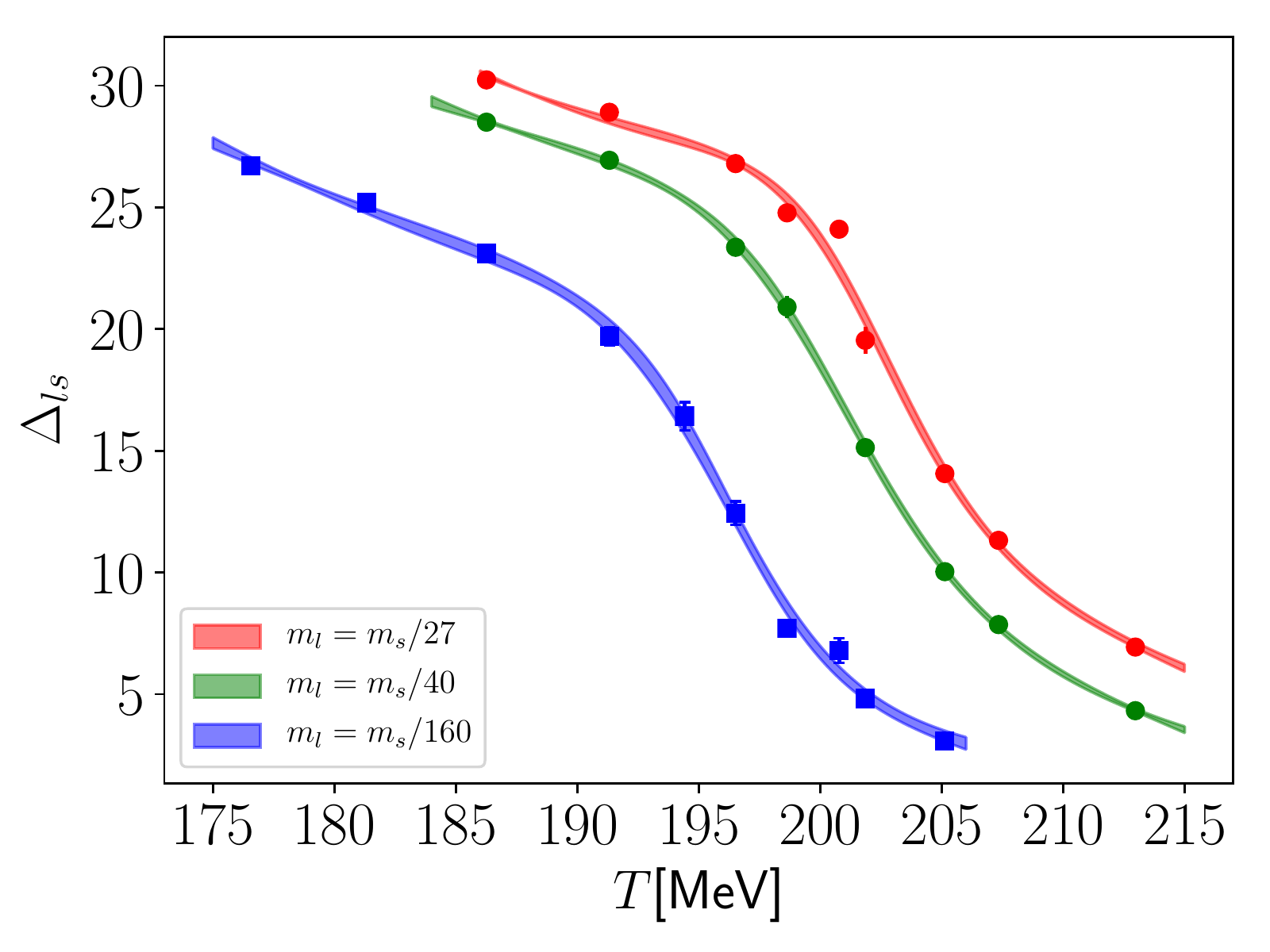}
        \includegraphics[scale=0.36]{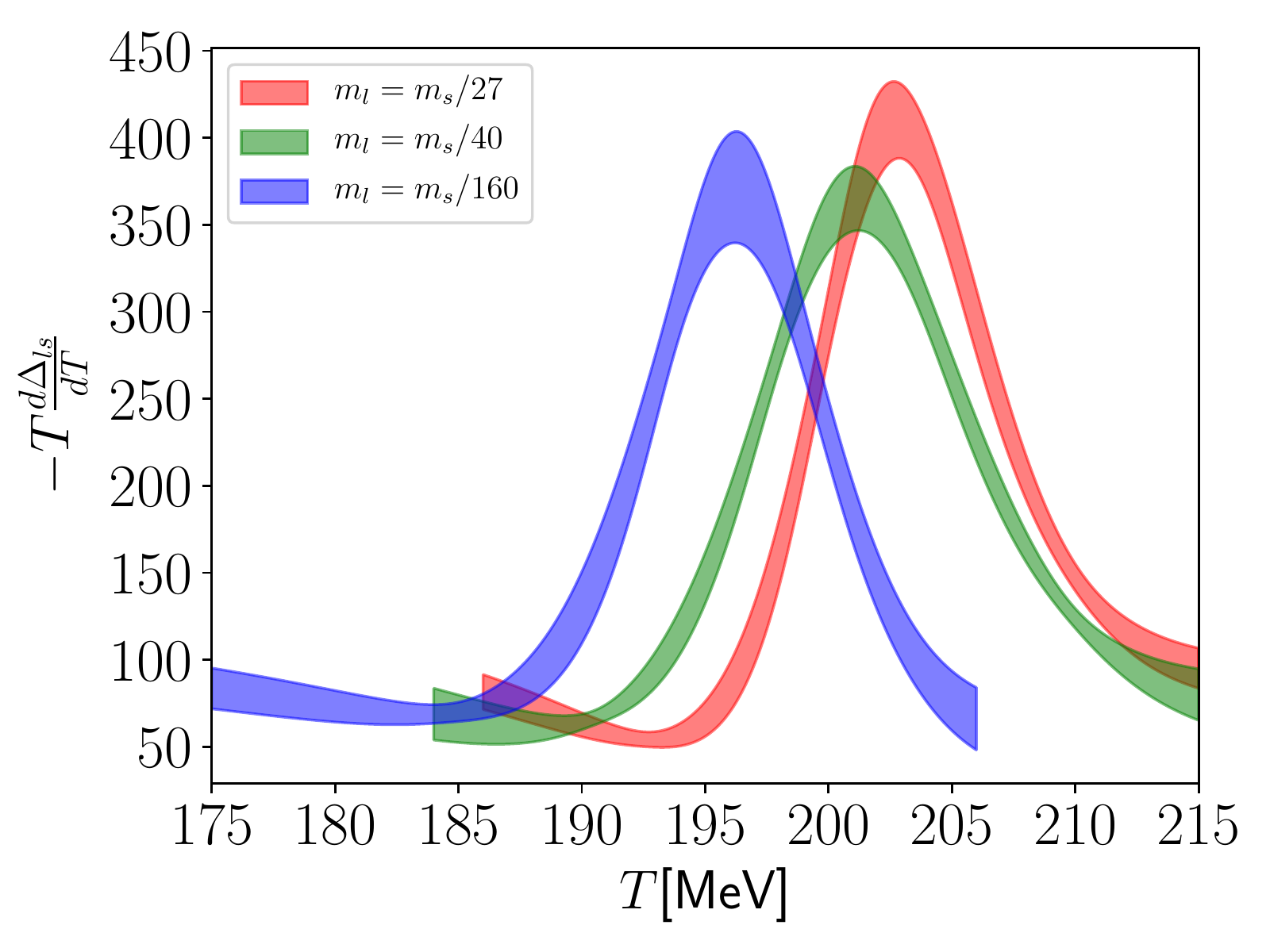}
        \caption{RGI chiral condensate (left) and its
temperature derivative (right) versus temperature and for
several values of the light quark masses. Shown are results
from simulations on lattices with spatial extent $N_\sigma =24$.}
\label{Rw_chiral}
\end{figure}

As discussed in Section 3 and shown in Fig.~\ref{contour}, the chiral 
condensate also may serve as an energy-like observable in the vicinity 
of the RW endpoint. The maximum in the temperature derivative of the 
chiral condensate, the mixed susceptibility, thus may equally well be
used to determine the temperature of the RW transition at fixed $m_l/m_s$. We 
show in Fig.~\ref{Rw_chiral} (left) results for the renormalization group invariant (RGI) chiral condensate 
\begin{equation}
\Delta_{ls} = \frac{m_s}{f_K^4} \left( \langle \bar{\psi}\psi \rangle_l - \frac{m_l}{m_s}
\langle \bar{\psi}\psi \rangle_s\right) \; ,
\end{equation}
where we used the kaon decay constant $f_K$ for normalization.
Its temperature derivative is shown Fig.~\ref{Rw_chiral} (right). The 
quark mass dependence of the peak location as well as
the weak dependence of the peak height on the quark mass 
is found to be in agreement with that
of the order parameter susceptibility  $\chi_h$.

\section{Conclusions}
\vspace{-0.2cm}
We have presented results from an ongoing study of the phase diagram
of (2+1)-flavor QCD in the
Roberge-Weiss plane on lattices with temporal extent $N_\tau=4$
using the HISQ action with light quark masses decreasing
from their physical value towards the chiral limit. Using a finite size
scaling analysis we found
that the endpoint of the $1^{st}$ order RW transition line remains 
$2^{nd}$ order at least down to light quark masses corresponding to Goldstone
pion masses of $55$~MeV. We have shown that the chiral condensate is 
sensitive to the RW transition and may serve as a energy-like operator
characterizing universal behavior close to the RW endpoint.

\section*{Acknowledgement}
\vspace{-0.2cm}
This work was supported in part through Contract No. DE-SC001270 with the
U.S. Department of Energy,
the Deutsche Forschungsgemeinschaft (DFG) through the grant CRC-TR 211 "Strong-interaction matter under extreme conditions"
and the grant 05P15PBCAA of the German Bundesministerium f\"ur Bildung und 
Forschung.

\vspace{-0.2cm}
\bibliographystyle{JHEP}
\bibliography{imag_mu}
\end{document}